\documentclass[nofootinbib]{revtex4}
\usepackage{graphicx}
\usepackage{latexsym}
\def\be{\begin{equation}}
\def\ee{\end{equation}}
\def\bea{\begin{eqnarray}}
\def\eea{\end{eqnarray}}

\begin{document}

\title{Cosmological CPT Violation, Baryo/Leptogenesis and CMB
Polarization}

\author{Mingzhe Li$^{1}$, Jun-Qing Xia$^{2}$, Hong Li$^{3}$
and Xinmin Zhang$^{2}$ }

\affiliation{${}^1$Institut f\"{u}r Theoretische Physik,
Philosophenweg 16, 69120 Heidelberg, Germany}
\affiliation{${}^2$Institute of High Energy Physics, Chinese
Academy of Science, P.O. Box 918-4, Beijing 100049, P. R. China}
\affiliation{${}^3$Department of Astronomy, School of Physics,
Peking University, Beijing, 100871, P. R. China}

\date{\today.}

\begin{abstract}
In this paper we study the cosmological $CPT$ violation and its
implications in baryo/leptogenesis and CMB polarization. We
propose specifically a variant of the models of gravitational
leptogenesis. By performing a global analysis with the Markov
Chain Monte Carlo (MCMC) method, we find the current CMB
polarization observations from the three-year WMAP (WMAP3) and the
2003 flight of BOOMERANG (B03) data provide a weak evidence for
our model. However to verify and especially exclude this type of
mechanism for baryo/leptogenesis with cosmological $CPT$ violation,
the future measurements on CMB polarization from PLANCK and CMBpol
are necessary.

\end{abstract}

\maketitle

\hskip 1.6cm PACS number(s): 98.80.Es, 98.80.Cq \vskip 0.4cm

\section{Introduction}

The $CPT$ symmetry which has been proved to be exact within the framework of
standard model of particle physics and Einstein gravity could be violated dynamically during
the evolution of the universe. To show it, consider a scalar boson
$\phi$ which
couples to a fermion current $J^\mu$, with the lagrangian given by
\be\label{baryon}
\mathcal{L}_{\rm int}=\frac{c}{M}\partial_{\mu}{\phi}J^{\mu}~,
\ee
where $c$ is a dimensionless constant and $M$ the cut-off scale of the
theory which could be the grand unification (GUT) or Planck scale.
The interaction in (\ref{baryon}) is $CPT$ conserved, however during the evolution of
the homogenous scalar field $\phi$ as the universe expands, $\dot \phi$
does not vanish which violates $CPT$ symmetry. This type of $CPT$-violation
occurs naturally in theory of dynamical dark energy and has two
interesting implications in particle physics and cosmology:

1) In models of quintessential baryo/leptogenesis
\cite{quin_baryogenesis,k-baryogenesis,DeFelice:2002ir}, the scalar field $\phi$
in (\ref{baryon}) is the dark energy scalar (quintessence, k-essence etc.).
The dynamics of the field $\phi$ leads to the current accelerating expansion
of the universe, meanwhile its interaction with the baryon
current $J^\mu_B$ (or $B-L$ current $J^\mu_{(B-L)}$) helps to produce the
baryon
number
asymmetry in thermal equilibrium. One of the features
of these models is a unified description of the present
accelerating expansion and the generation of the matter and
antimatter asymmetry of our universe. Furthermore differing from
the original proposal for spontaneous baryogenesis by Cohen and
Kaplan \cite{spon_baryogenesis} since the quintessence scalar
has been existing up to present epoch the corresponding $CPT$-violation
could be
tested in laboratory experiments and cosmology. Along this line,
the gravitational baryo/leptogenesis \cite{steinhardt,lihong} has
been proposed in which a function of curvature scalar $R$ replaces
the $\phi$ field in (\ref{baryon}).

2)Replacing the fermion current in (\ref{baryon}) by
a Chern-Simons current $A_{\nu}\widetilde{F}^{\mu\nu}$, where $A_{\nu}$ is
the electromagnetic vector potential,
$F_{\mu\nu}=\nabla_{\mu}A_{\nu}-\nabla_{\nu}A_{\mu}$ is the strength
tensor and
$\widetilde{F}^{\mu\nu}=1/2\epsilon^{\mu\nu\rho\sigma} F_{\rho\sigma}$
is its dual,
will modify the
Maxwell equations and gives rise to interesting and observable phenomena
in the photon
sector.
The Chern-Simons term can lead to the rotations of the polarization vectors of
photons when propagating over the
cosmological distance \cite{jackiw}, dubbed as
``cosmological birefringence". The change in the
position angle of the polarization plane $\Delta\chi$,
can be obtained by observing polarized radiation from distant sources
such as radio
galaxies,
 quasars \cite{jackiw,field,CF97,carroll} and
the cosmic microwave background (CMB) \cite{kamionkowski}.
Recently, in a model independent way, three of us (M. Li, J.-Q.
Xia, and X. Zhang) with B. Feng and X. Chen in Ref.
\cite{cptv_boomrang} found that a nonzero rotation angle of the
photons is mildly favored by the current CMB polarization data
from the three year Wilkinson Microwave Anisotropy Probe (WMAP3)
observations
 \cite{Spergel:2006hy,Page:2006hz,Hinshaw:2006,Jarosik:2006,WMAP3IE}
and the January 2003 Antarctic flight of BOOMERANG (Hereafter
B03)\cite{B03,B03EE,TCGC}(see also Ref.\cite{Liu:2006uh}). This is
a signal in some sense of the cosmological $CPT$-violation
mentioned above.  It would be interesting to ask if the
measurements on CMB polarization mentioned here serve as a test of
the mechanism of baryo/leptogenesis above?

In this paper we discuss the possible connection between the two
phenomena above associated with the cosmological $CPT$ violation and
study the possibility of testing the baryo/leptogenesis with CMB
polarization. We in this paper specifically propose a class of
models which give rise to enough amount of baryon number asymmetry
required and predict sizable rotations of CMB photons observable
in the CMB polarization measurements. The paper is organized as
follows. In section II we present the model for leptogenesis.  In
section III we study the $CPT$ violating effects of
our model on the
electromagnetic theory. We use the geometrical optics
approximation to get the
 equation describing the rotations of polarizations of photons.
In section IV the model is confronted with the data from
 CMB experiments.
Section V is our conclusion.

\section{Our Model of Leptogenesis}

We search for variants of the models of the quintessential or
gravitational baryo/leptogenesis and study their implications in CMB
polarization. To begin our discussions we start with gravitational
baryo/leptogenesis. For this model the dark energy is given by the
cosmological constant. Compared with the quintessential baryo/leptogenesis
where the potential needs to be specified this will simplify our
study, however the results obtained can be easily generalized into the cases
for the quintessential baryo/leptogenesis.

The model under investigation is a variant of the model proposed
in Ref.\cite{lihong} with the lagrangian \be\label{model}
\mathcal{L}_{\rm int}= -c~ \partial_{\mu} \ln R J^{\mu}_{i}~, \ee
where $J^{\mu}_{i}$ is the fermion current.
To study
the possible connection between the CPT violations in gravitational
baryo/leptogenesis and CMB polarization \footnote{The possible connection
between the CPT violations in quintessential baryogenesis and the
cosmological birefringence has been considered in
 Ref.\cite{DeFelice:2002ir}.}
$J^{\mu}_i$ is required to satisfy
i) not orthogonal to the baryon or $B-L$ current; and ii)
anomalous with respect to the electromagnetic interaction.
The
first condition above guarantees the generation of baryon number
asymmetry and the second one makes it possible to be connected
with the CMB polarization.
There are various possibilities for $J^{\mu}_i$ required.
One of the examples is $J^{\mu}_i$ being the Peccei-Quinn
current \cite{PRL}. Here for the specific discussion we take
$J^{\mu}_i = J^{\mu}_{(B-L)_{\rm L}}$, the left-handed part of
$B-L$ current. Formally this current can be decomposed into the
combination of a $B-L$ current and an axial $B-L$ current, namely
$J^{\mu}_{(B-L)_{\rm L}} = (1/2) J^{\mu}_{(B-L)} - (1/2)J^{5
\mu}_{(B-L)}$. We will show below that the contribution of
$J^{\mu}_{(B-L)}$ to the generation of the baryon number asymmetry
will dominate over $J^{5 \mu}_{(B-L)}$\footnote{$J^{5
\mu}_{(B-L)}$ is violated by the Yukawa couplings, which give rise
to a lower decoupling temperature $T_D \sim 100$ GeV.}, however
$J^{5 \mu}_{(B-L)}$ plays an essential role in connection to the
effects in CMB polarization.

Following the calculations in Ref. \cite{lihong}, we obtain the final
baryon number asymmetry \bea\label{result} \frac{n_B}{s} \simeq
\frac{2.52}{2}g_b g_{\ast s}^{-1/2}\frac{T_D}{m_{pl}}\simeq 0.24c\frac{T_D}{m_{pl}}~, \eea
where $g_b=2$ is the number of degrees of freedom of baryons,
$g_{\ast s}$ characterizes the total degrees of
freedom of all the relativistic particles in the universe,
and its value is about $106.75$ at the epoch when the temperature is around
$O(100)$GeV or higher.
$T_D$ is the decoupling
temperature determined by the $B-L$ violating interaction, which
in Ref.\cite{lihong} is given by
\begin{eqnarray}\label{lepvio}
{\cal L}_{\not L} = \frac{2} { F } l_L l_L \Phi \Phi +{\rm H.c.}~,
\end{eqnarray}
where $F$ is a scale of new physics beyond the Standard Model of
particle physics which generates the $B-L$ violations, $\Phi$ is
the Higgs doublet and $l_L$ the left-handed lepton doublet. When
the Higgs field gets a vacuum expectation value $< \Phi > \sim v
$, the left-handed neutrino receives a Majorana mass $m_\nu \sim
\frac{v^2}{F}$ .

In the early universe the lepton number violating rate induced by
the interaction in (\ref{lepvio}) is \cite{sarkar}
\begin{eqnarray}
  \Gamma_{\not L} \sim
    0.04~ \frac{T^3}{ F^2 }~.
\end{eqnarray}
Since $\Gamma_{\not L}$ is proportional to $T^3$, for a given $F$,
namely the neutrino mass, $B-L$ violation will be more efficient at
high temperature than at low temperature. Requiring this rate be
larger than the Universe expansion rate $H\simeq 1.66
g_{\ast}^{1/2}T^2/ m_{pl}$ until the temperature $T_D$, we obtain a
$T_D$-dependent lower limit on the neutrino mass:
\begin{eqnarray}\label{neutrinomass}
   \sum_i m_i^2  = ( 0.2 ~{\rm eV} ( { \frac{10^{12}~{\rm GeV}}{T_D}
})^{1/2})^2 .
\end{eqnarray}
The experimental bounds on the neutrino masses come from the neutrino
oscillation experiments and the cosmological tests. The
atmospheric and solar neutrino oscillation experiments
give\cite{atmospheric,solar}: \bea & & \Delta
m_{atm}^2=(2.6\pm 0.4)\times 10^{-3} {\rm eV}^2~,\\
& & \Delta m_{sol}^2 \simeq (8.0^{+0.6}_{-0.4})\times 10^{-5} {\rm
eV}^2~. \eea The cosmological tests provide the limits on the sum
of the three neutrino masses, $\Sigma\equiv \sum_i m_i$. The
analysis of WMAP3 \cite{Spergel:2006hy} and SDSS \cite{Tegmark}
show the constraints: $\Sigma<0.68$ eV and $\Sigma<0.94$ eV
respectively.

For the case of normal hierarchy neutrino masses, $m_3\gg m_2,~m_1
$, one has \be
 m_3^2-m_2^2=\Delta m_{atm}^2~,~~~ m_2^2-m_1^2=\Delta m_{sol}^2~,
\ee and \be \sum_i m_i^2\simeq m_3^2\gtrsim \Delta m_{atm}^2. \ee
We can see from Eq. (\ref{neutrinomass}) that this requires the
decoupling temperature $T_D\lesssim 1.6\times 10^{13}$ GeV. For
neutrino masses with inverted hierarchy, $m_2\sim m_1\gg m_3$, we
get \be
 m_2^2-m_3^2=\Delta m_{atm}^2~,~~~ m_2^2-m_1^2=\Delta m_{sol}^2~,
\ee and
 \be \sum_i m_i^2\simeq 2m_2^2\gtrsim 2\Delta m_{atm}^2 , \ee
which constrains the decoupling temperature as $T_D\lesssim
8\times10^{12}$ GeV. If three neutrino masses are approximately
degenerated, ~{\rm i.e.}~, $m_1 \sim m_2 \sim m_3\sim {\bar m}$,
one has $\Sigma = 3 {\bar m}$ and $\sum_i m_i^2 \simeq
\Sigma^2/3$. In this case, the WMAP3 and SDSS data require $T_D$
to be larger than $2.6\times 10^{11}$ GeV and $1.4\times 10^{11}$
GeV respectively. So, for a rather conservative estimate, we
consider $T_D$ in the range of $1.4\times 10^{11} {\rm
GeV}\lesssim T_D \lesssim 1.6\times 10^{13} {\rm GeV}$. Combined
with Eq. (\ref{result}) and the observational limit for the
baryon/photon ratio \cite{pdg}, $n_B/n_{\gamma}=7.04
n_B/s=(6.15\pm 0.25)\times 10^{-10}$, we find that a successful
leptogenesis requires the coupling constant $c$ should to be
larger than $2.55\times 10^{-4}$ (at $2\sigma$).

\section{Modified Electromagnetic Theory and Cosmological Birefringence}

The current $J^{\mu}_{(B-L)_{\rm L}}$ is anomalous under the
electromagnetic interaction. Using the notation of \cite{Peskin}
we obtain

\be
\partial_{\mu}J^{\mu}_{(B-L)_{\rm L}} \sim
-\frac{e^2}{12\pi^2}F_{\mu\nu}\widetilde{F}^{\mu\nu}=
-\frac{\alpha_{em}}{3\pi}F_{\mu\nu}\widetilde{F}^{\mu\nu}~, \ee
where $\alpha_{em}=e^2/4\pi=1/137$ is the electromagnetic fine structure
constant. So, by an integration in part one can see that the
interaction in (\ref{model}) induces a
Chern-Simons term, \be \mathcal{L}_{\rm int}\sim
-\frac{c\alpha_{em}}{3\pi}\ln R F_{\mu\nu}\widetilde{F}^{\mu\nu}
\sim \frac{2c\alpha_{em}}{3\pi}\partial_{\mu}\ln R
A_{\nu}\widetilde{F}^{\mu\nu}~. \ee Hence the vacuum Maxwell
equations are modified as \be \nabla_{\mu}F^{\mu\nu}=\delta
\partial_{\mu}\ln R \widetilde{F}^{\mu\nu}~, \ee or in terms of
the vector field $A_{\mu}$, \be\label{maxwell}
\nabla_{\mu}(\nabla^{\mu}A^{\nu}-\nabla^{\nu}A^{\mu})={\delta\over
2} \partial_{\mu}\ln R
\epsilon^{\mu\nu\rho\sigma}(\nabla_{\rho}A_{\sigma}-\nabla_{\sigma}A_{\rho})~,
\ee in above we have defined \be \delta \equiv -\frac{4
\alpha_{em}}{3\pi}c~. \ee The equation (\ref{maxwell}) contains
gauge freedom. Usually we impose a gauge condition to get rid of
it. In this paper, we impose Lorentz gauge
\be\label{gauge} \nabla_{\mu}A^{\mu}=0~. \ee Under this gauge, Eq. (\ref{maxwell}) becomes \be
\nabla_{\mu}\nabla^{\mu}A^{\nu}+R^{\nu}_{\mu}A^{\mu}={\delta\over
2} \partial_{\mu}\ln R
\epsilon^{\mu\nu\rho\sigma}(\nabla_{\rho}A_{\sigma}-\nabla_{\sigma}A_{\rho})~,
\ee where the Ricci tensor $R^{\nu}_{\mu}$ appeared due to the
commutation of covariant derivatives acting on a vector field.

In the cases where the scale of variation of electromagnetic field
 is much smaller than that we are interested in as we do in this paper,
the geometrical optics approximation is applicable. In such an approximation,
the solutions to the Maxwell equations are
expected to be in the form \cite{Misner:1974qy},
\be\label{geometry}
A^{\mu}={\rm Re}[(a^{\mu}+\epsilon b^{\mu}+\epsilon^2 c^{\mu}+...)e^{iS/\epsilon}]~,
\ee
with $\epsilon$ a small coefficient. It means that the
phase $S/\epsilon$ varies much faster than the
amplitude does.
With the help of it, one can easily see that
the gauge condition (\ref{gauge}) implies
\be\label{orth}
k_{\mu}a^{\mu}=0~,
\ee
where the wave vector $k_{\mu}\equiv \nabla_{\mu}S$ is
orthogonal to the surfaces of constant phase and represents the direction
which photons
travel along. We can write the vector $a^{\mu}$ as the product of a
scalar amplitude $A$
and a normalized polarization vector $\varepsilon^{\mu}$,
\be\label{split}
a^{\mu}=A\varepsilon^{\mu}~,
\ee
with
\be\label{normalization}
\varepsilon_{\mu}\varepsilon^{\mu}=1~.
\ee
We can see that under the Lorentz gauge, the wave vector $k_{\mu}$ is
orthogonal to the polarization vector $\varepsilon^{\mu}$.
Substituting the equation (\ref{geometry}) into the modified Maxwell equation (\ref{maxwell}) and neglecting the Ricci tensor
term yield
\bea
& &\nabla_{\mu}\nabla^{\mu}(a^{\nu}+\epsilon b^{\nu}+...)+\frac{2i}{\epsilon}k^{\mu}
\nabla_{\mu}(a^{\nu}+\epsilon b^{\nu}+...)+\frac{i}{\epsilon}(\nabla_{\mu}k^{\mu})
(a^{\nu}+\epsilon b^{\nu}+...)-\frac{1}{\epsilon^2}k_{\mu}k^{\mu}
(a^{\nu}+\epsilon b^{\nu}+...)\nonumber\\
& &={\delta\over 2} \partial_{\mu}\ln R \epsilon^{\mu\nu\rho\sigma}
\{[\nabla_{\rho}(a_{\sigma}+\epsilon b_{\sigma}+...)-\nabla_{\sigma}(a_{\rho}+\epsilon b_{\rho}+...)]
+\frac{i}{\epsilon}[k_{\rho}(a_{\sigma}+
\epsilon b_{\sigma}+...)-k_{\sigma}(a_{\rho}+\epsilon b_{\rho}+...)] .\}
\eea
Collecting respectively the terms proportional to $1/\epsilon^2$ and
$1/\epsilon$ in both sides of above equation, we have
\be\label{null}
k_{\mu}k^{\mu}=0~,
\ee
and
\be\label{main}
k^{\mu}\nabla_{\mu}a^{\nu}+\frac{1}{2}\nabla_{\mu}k^{\mu}a^{\nu}=\frac{\delta}{4}\partial_{\mu}\ln R\epsilon^{\mu\nu\rho\sigma}
(k_{\rho}a_{\sigma}-k_{\sigma}a_{\rho})~.
\ee
The equation (\ref{null}) indicates that photons propagate along the null geodesics, which are unaffected by the
Chern-Simons term.
This can be seen from its differentiation,
\be\label{null2}
0=\nabla_{\nu}(k_{\mu}k^{\mu})=2\nabla^{\mu}S\nabla_{\nu}\nabla_{\mu}S=2\nabla^{\mu}S\nabla_{\mu}\nabla_{\nu}S\nonumber\\
=2k^{\mu}\nabla_{\mu}k_{\nu}~.
\ee
We can define the affine parameter $\lambda$ which measures the distance along the light-ray,
\be
k^{\mu}\equiv \frac{dx^{\mu}}{d\lambda}~.
\ee
So, from Eq. (\ref{null2}) we get the geodesic equation
\be
\frac{dx^{\mu}}{d\lambda}\nabla_{\mu}k_{\nu}=0~.
\ee
The effect of Chern-Simons term appears in the equation (\ref{main}). Multiplying
Eq. (\ref{main}) with $a_{\nu}$ and using the product (\ref{split}) and
the normalization (\ref{normalization}) give,
\be
\nabla_{\mu}k^{\mu}=-2k_{\mu}\nabla^{\mu}\ln A~.
\ee
In terms of it, Eq. (\ref{main}) gives the following equations describing how the polarizaiton vector changes,
\be\label{main2}
k^{\mu}\nabla_{\mu}\varepsilon^{\nu}=\frac{\delta}{4}\partial_{\mu}\ln R\epsilon^{\mu\nu\rho\sigma}
(k_{\rho}\varepsilon_{\sigma}-k_{\sigma}\varepsilon_{\rho})~.
\ee
We see that the Chern-Simons term makes
 $k^{\mu}\nabla_{\mu}\varepsilon^{\nu}$ not vanished. This means that the polarization vector $\varepsilon^{\nu}$ is not
 parallelly transported along the light-ray. It rotates as the photon propagates in spacetime.

We consider here the spacetime described by spatially flat
Friedmann-Robertson-Walker (FRW) metric
\be
ds^2=a^2(d\eta^2-\delta_{ij}dx^idx^j)~,
\ee
and the curvature scalar $R$ is homogeneous, i.e., only varies with time.
The null condition (\ref{null}) is
\be
(k^0)^2-k^ik^i=0~.
\ee
We assume that photons propagate along the positive direction of x axis, i.e., $k^{\mu}=(k^0,~ k^1,~ 0,~ 0)$
and $k^1/k^0>0$. Above equation implies $k^1= k^0$. Gauge invariance
guarantees that the polarization vector of the photon
has only two independent components which are orthogonal to
the propagating direction. So,
we are only interested in how the components of the polarization
vector, $\varepsilon^2$
and $\varepsilon^3$, change. They satisfy the following equations,
\be
k^{\mu}\nabla_{\mu}\varepsilon^i=\frac{\delta}{2}\frac{\dot R}{R} \epsilon^{0ijk}k_j\varepsilon_k
=\frac{\delta}{2}\frac{\dot R}{R} \frac{e^{0ijk}}{\sqrt{-{\rm det}|g|}}k_j\varepsilon_k
=\frac{\delta}{2}\frac{\dot R}{R} \frac{e^{ijk}}{a^4}k_j\varepsilon_k=\frac{\delta}{2}\frac{\dot R}{R} e^{ijk}
k^j\varepsilon^k=-\frac{\delta}{2}\frac{\dot R}{R} e^{1ik}k^1\varepsilon^k~, ~~~{\rm with}~i=2,~3~,
\ee
where the dot means derivative with respect to the conformal time $\eta$. We have used $\epsilon^{\mu\nu\rho\sigma}=e^{\mu\nu\rho\sigma}/\sqrt{-{\rm det}|g|}$ and the components of the
total antisymmetric tensor density $e^{\mu\nu\rho\sigma}$ are set to be $e^{0123}=1$ and so on.
$e^{ijk}$ is three dimensional total antisymmetric tensor density with $e^{123}=1$. So we can match that $e^{0ijk}=e^{ijk}$.
 With the helps of $k^{\mu}=dx^{\mu}/d\lambda$ and the
Christoffel symbols,
\be
\Gamma^i_{0j}=\Gamma^i_{j0}=\mathcal{H}\delta^i_j~,~~~\Gamma^i_{00}=\Gamma^i_{jk}=0~,
\ee
we get,
\bea
& &\frac{d\varepsilon^2}{d\lambda}+\mathcal{H}k^0\varepsilon^2=-\frac{\delta}{2}\frac{\dot R}{R}k^1\varepsilon^3~,\\
& &\frac{d\varepsilon^3}{d\lambda}+\mathcal{H}k^0\varepsilon^3=
\frac{\delta}{2}\frac{\dot R}{R}k^1\varepsilon^2~,
\eea
where we have defined $\mathcal{H}\equiv \dot a/a$.
Furthermore, using $k^1=k^0=d\eta/d\lambda$ yields,
\bea
& &\frac{d}{d \ln R}(a\varepsilon^2)=-\frac{\delta}{2}a\varepsilon^3~,\\
& &\frac{d}{d \ln
R}(a\varepsilon^3)=\frac{\delta}{2}a\varepsilon^2~. \eea It is
easy to find that the polarization angle is \be
\alpha\equiv\arctan{(\frac{\varepsilon^3}{\varepsilon^2})}=\frac{\delta}{2}\ln
R+{\rm constant}~. \ee Hence, for a source at a redshift $z$, the polarization
angle is rotated by \be\label{kmu} \Delta
\alpha=\frac{\delta}{2}\ln{\left(\frac{R(0)}{R(z)}\right)}~. \ee
As we know, a vector rotated by an angle $\Delta\alpha$ in a fixed
coordinates frame is equivalent to a fixed vector observed in a
coordinates frame which is rotated by $-\Delta\alpha$. So, with
the notion of coordinates frame rotation, the rotation angle is
\be\label{kmu2} \Delta\chi=-\Delta\alpha=\frac{\delta}{2} \ln
\left(\frac{R(z)}{R(0)}\right)~. \ee

\begin{figure}[htbp]
\begin{center}
\includegraphics[scale=1.5]{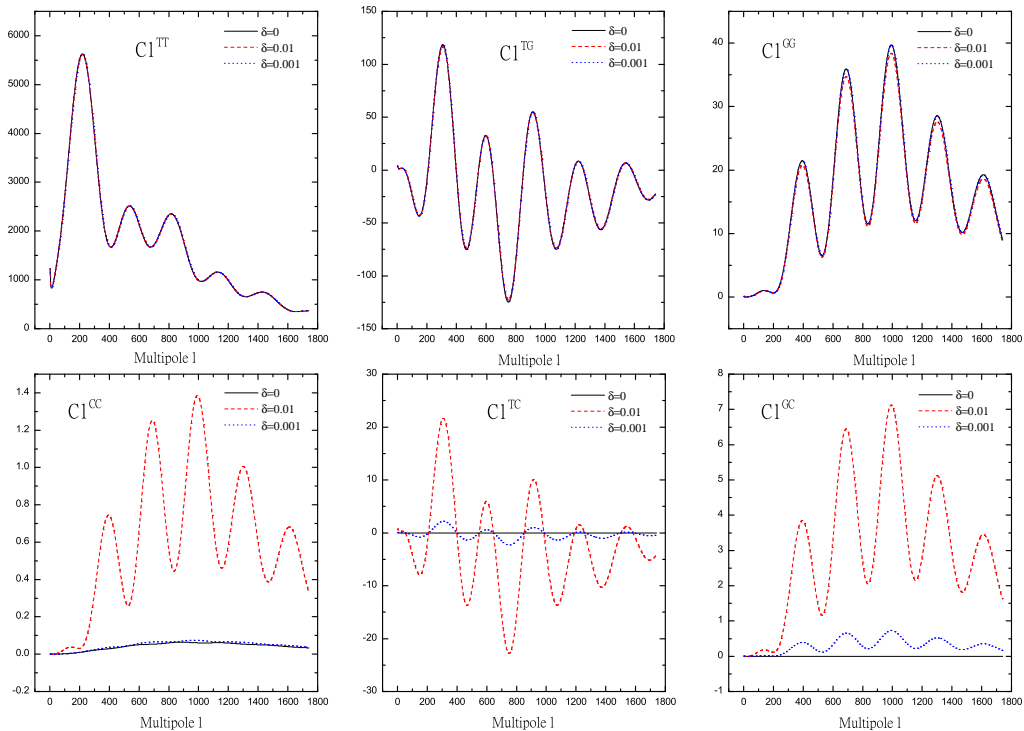}
\caption{ (color online). The effects of the rotation angle
$\Delta\chi$ in Eq. (\ref{kmu2}) on the power spectra of TT, TG,
GG, CC, TC and GC. The black solid line is for the case of
$\delta=0$, the red dashed line is for $\delta=0.01$ and the blue
dotted line is for $\delta=0.001$. \label{sig}}
\end{center}
\end{figure}

\section{CMB polarization}

Now we consider the constraints on the parameter $\delta$ in
(\ref{kmu2}) from the current CMB data. The cosmic microwave background can
be
characterized by the temperature and polarization completely in
each direction on the sky. Usually we describe the photon that
come to us in terms of the Stokes parameters $I$, $Q$, $U$ and
$V$\cite{stoks_parametr}. $I$ and $V$ describe physical
observables and are independent of the choice of the coordinate
system. However, $Q$ and $U$ which characterize the orthogonal
modes of the linear polarization depend on the axes where the
linear polarization is defined. $Q$ and $U$ can be decomposed into
a gradient-like (G) and a curl-like (C) components \cite{HW97}. If
the $\delta$ term is absent
the TC and GC cross correlations power spectra vanish.
However, with the presence of cosmological birefringence, the
polarization vector of each photon is rotated by an angle $\Delta
\chi$, which can give rise to non-zero TC and GC correlations,
even though they are zero at the last scattering surface. With a
prime we give the rotated quantities \cite{kamionkowski}
\begin{equation}
\label{TC} C_l'^{TC}= C_l^{TG}\sin 2 \Delta \chi~
\end{equation}
and \cite{grav_cmb}\begin{equation} \label{GC} C_l'^{GC}=
\frac{1}{2}(C_l^{GG}- C_l^{CC})\sin 4 \Delta \chi ~.
\end{equation}
Also, the original TG, GG and CC spectra are modified as following:
\begin{eqnarray}
\label{TG}
C_l'^{TG} &=& C_l^{TG}\cos 2 \Delta \chi ~,\\
\label{GG}
C_l'^{GG} &=& C_l^{GG}\cos^2 2 \Delta \chi + C_l^{CC}\sin^2 2 \Delta \chi~,\\
\label{CC} C_l'^{CC} &=& C_l^{CC}\cos^2 2 \Delta \chi +
C_l^{GG}\sin^2 2\Delta \chi~ ~.
 \end{eqnarray}
 The CMB temperature power spectrum remains unchanged.

With the notation given above we in fig.\ref{sig} illustrate the
effects of the rotation angle $\Delta\chi$ in Eq. (\ref{kmu2}) for
three cases $\delta=0$, $\delta=0.01$ and $\delta=0.001$. The
basic cosmological parameters we choose for these plots are given
below and consistent with the current result of
WMAP3\cite{Spergel:2006hy}:
\begin{equation}
(\Omega_b h^2, \Omega_m h^2, \tau, H_0, n_s, A_s) = (0.0223,
0.1265, 0.088, 73.5, 0.951, 2.37\times10^{-9})~.
\end{equation}
One can see from fig.\ref{sig} that the C-mode is very sensitive
to $\delta$. So we expect the TC and GC data
constrain $\delta$ efficiently.

To determine the parameter $\delta$,
we make a global fitting to the CMB data
with the publicly available Markov Chain Monte Carlo package
cosmomc\cite{Lewis:2002ah,IEMCMC}. During the calculations, we
modified the programs by adding a new free parameter $\delta$ in
Eq.(\ref{kmu2}) to allow the rotation of the power spectra
given above. We sample the following 8 dimensional set of
cosmological parameters:
\begin{equation}\label{para}
    \textbf{p}\equiv(\omega_{b},\omega_{c},\Theta_S,\tau,\delta,n_{s},r,\log[10^{10}A_{s}])
\end{equation}
where $\omega_{b}=\Omega_{b}h^{2}$ and $\omega_{c}=\Omega_{c}h^{2}$
are the physical baryon and cold dark matter densities relative to
critical density, $\Theta_S$ (multiplied by $100$) is the ratio of
the sound horizon and angular diameter distance, $\tau$ is the
optical depth, $A_{s}$ is defined as the amplitude of initial power
spectrum and $n_{s}$ measures the spectral index. Basing on the
Bayesian analysis, we vary the above 8 parameters fitting to the
observational data (such as WMAP3 and B03 data) with the MCMC
method. Throughout, we assume a flat universe and
take the weak priors as: 
$0.01<\tau<0.8, 0.5<n_{s}<1.5$, a cosmic age tophat prior as 10
Gyr$<t_{0}<$20 Gyr. Furthermore, we make use of the HST measurement
of the Hubble parameter $H_0 = 100h \quad \text{km s}^{-1}
\text{Mpc}^{-1}$ \cite{freedman} by multiplying the likelihood by a
Gaussian likelihood function centered around $h=0.72$ and with a
standard deviation $\sigma = 0.08$. We impose a weak Gaussian prior
on the baryon and density $\Omega_b h^2 = 0.022 \pm 0.002$ (1
$\sigma$) from Big Bang nucleosynthesis\cite{bbn}.

In our calculations we have taken the total likelihood to be the
products of the separate likelihoods of WMAP3 and B03. Alternatively
defining $\chi^2 = -2 \log {\bf \cal{L}}$, we get \be \chi^2_{total}
= \chi^2_{WMAP3}+ \chi^2_{B03}.\ee
In the computation of WMAP3 data
 we have included the cross correlations power spectra of TT, GG, TG
 and CC, while for
 the B03 data we added two more datasets: the spectra GC and
TC, which give rise to the direct $CPT$-violation signal.

\begin{figure}[htbp]
\begin{center}
\includegraphics[scale=0.55]{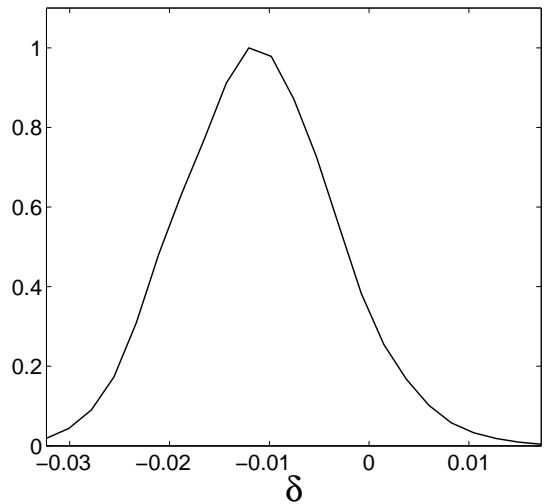}
\caption{ (color online). One dimensional constraints on the
parameter $\delta$ from WMAP3 and B03. \label{1demension}}
\end{center}
\end{figure}

In the Fig.\ref{1demension} we plot the one dimensional constraint
on the parameter $\delta$ from the current CMB polarization
experiments. The numerical results give that the $1, 2 \sigma$
limits: $\delta=-0.011^{+0.0074+0.0157}_{-0.0079-0.0142}$. This
shows a weak preference for a non-zero $\delta$. But at $2 \sigma$
$\delta$ is consistent with zero, which put a limit $| \delta | <
10^{-2}$.

\section{Conclusions}
In this paper we have studied the cosmological $CPT$-violation and
its implications in cosmology. We proposed a specific model where
baryo/leptogenesis is connected to the CMB polarization. Our model
can generate the baryon number asymmetry required and
predict a sizable effect on CMB polarization observable at the
current measurements. Using the current CMB data, we have performed a global fitting to
determine the rotation angle in $\Lambda$CDM cosmology.  We
found the current data mildly favor a non-vanishing $\delta$ which
provides a weak evidence for the cosmological $CPT$-violation.
However to verify and especially exclude this class of models of
baryo/leptogenesis, more precision measurements on the CMB
polarization from PLANCK and CMBpol are required\cite{grav_cmb}.

\textbf{Acknowledgement}: We acknowledge the use of the Legacy
Archive for Microwave Background Data Analysis (LAMBDA). Support
for LAMBDA is provided by the NASA Office of Space Science. We
have performed our numerical analysis on the Shanghai
Supercomputer Center(SSC). We thank Pei-Hong Gu, Xiao-Jun Bi, Yi-Fu Cai, Bo
Feng and Gong-Bo Zhao for helpful discussions. The author M.L. is grateful to the
support from Alexander von Humboldt Foundation. This work is
supported in part by National Natural Science Foundation of China
under Grant Nos. 90303004, 10533010 and 19925523.

{}

\end{document}